\documentclass[achemso,twocolumn,amsmath,amssymb,longbibliography]{revtex4-1}
\usepackage{graphicx,epsf}
\usepackage{dcolumn}
\usepackage{bm}

\begin{document}


\def\comment#1{{\large\textsl{#1}}}
\def\eq#1{{Eq.~(\ref{eq:#1})}}
\def\fig#1{{Fig.~\ref{fig:#1}}}
\def\sec#1{{Sec.~\ref{sec:#1}}}
\def\Ref#1{{Ref.~\onlinecite{#1}}}  
\def\tab#1{{Table~\ref{tab:#1}}}

\title{Thermodynamics of the Flexible Metal-Organic Framework Material MIL-53(Cr)
From First Principles}

\author{Eric Cockayne}

\affiliation{Materials Measurement Science Division, 
Material Measurement Laboratory, 
National Institute of Standards and Technology, 
Gaithersburg, Maryland 20899 USA}


\begin{abstract}

 We use first-principles density functional theory total energy and linear response phonon calculations 
to compute the Helmholtz and Gibbs free energy as a function of temperature, pressure, and cell volume  
in the flexible metal-organic framework material MIL-53(Cr) within the quasiharmonic approximation.  
GGA and metaGGA calculations were performed, each including empirical van der Waals (vdW) forces 
under the D2, D3, or D3(BJ) parameterizations.  At all temperatures up to 500 K and pressures
from -30 MPa to 30 MPa,  two minima in the free energy versus volume are found, corresponding to the narrow pore ($np$) and 
large pore ($lp$) structures.   Critical positive and negative pressures are identified,
beyond which there is only one free energy minimum.  While all results overestimated the stability of the $np$ phase relative 
to the $lp$ phase, the best overall agreement with experiment is found for the metaGGA PBEsol+RTPSS+U+J 
approach with D3 or D3(BJ) vdW forces.  For these parameterizations, the calculated 
free energy barrier for the $np$-$lp$ transition is only 3 to 6 kJ per mole of
Cr$_4$(OH)$_4$(C$_8$H$_4$O$_4$)$_4$.  

\end{abstract}

\maketitle



\begin{figure}
\includegraphics[width=85mm]{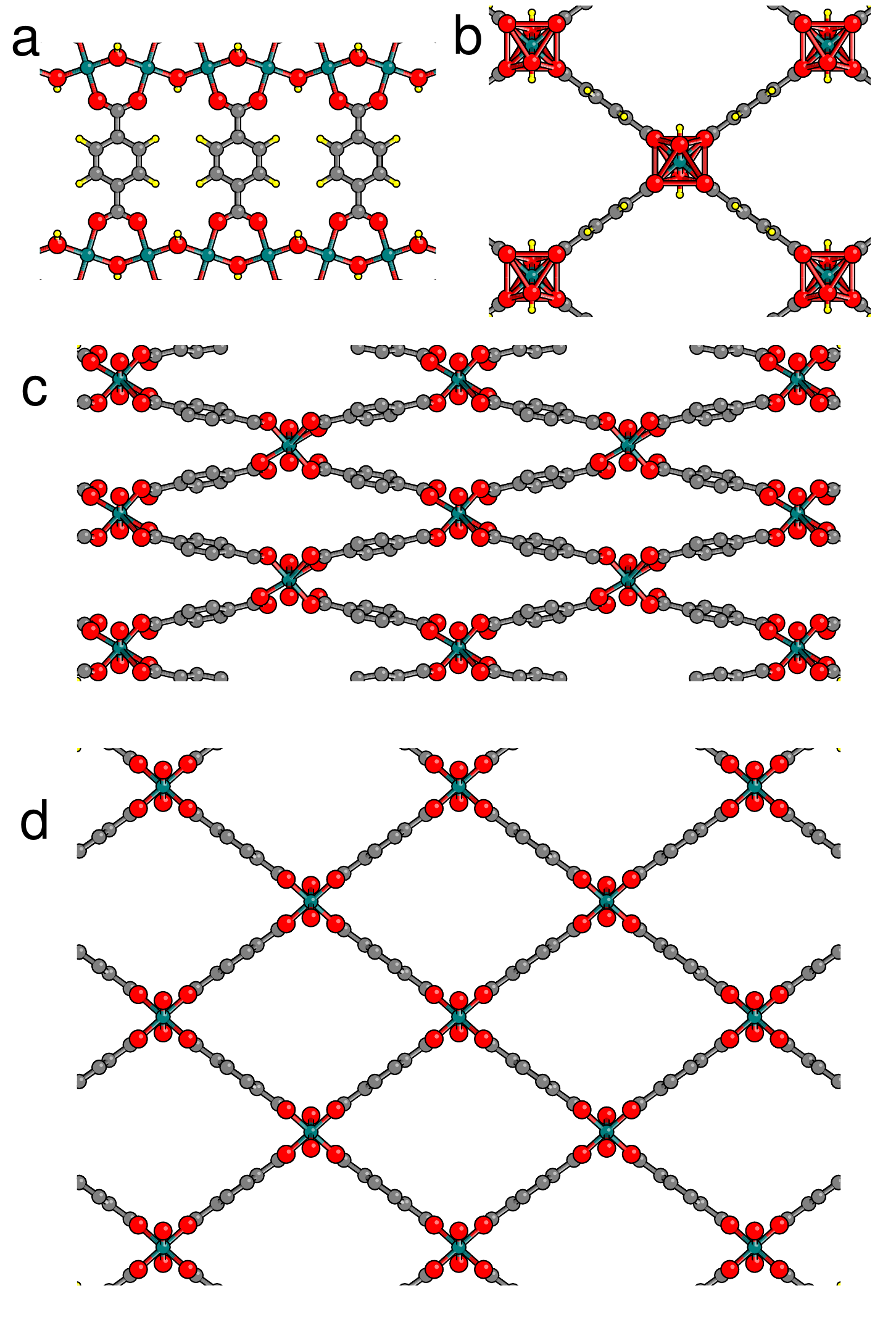}
\caption{Structure of MIL-53(Cr). Cr atoms green, O red,
C gray, and H yellow. (a) bdc linkers joining zigzag Cr-OH-Cr-$\dots$ chains. 
(b) Each zigzag chain is coordinated with four neighboring chains; 
each Cr is octahedrally coordinated with six O. (c) Narrow pore ($np$) phase
showing bdc rotations. (d) Large pore ($lp$) phase.  In (c) and (d), the
H are not shown.}
\label{fig:mil53x}
\end{figure}

Microporous flexible metal-organic framework materials 
are fascinating both from a fundamental point of
view and for their numerous potential applications such as
gas storage, gas separation, sensors, drug delivery, etc.\cite{Ferey09,
Alhamami14,Schneemann14,Coudert15,Ferey16}
A well-studied example is the MIL-53 
family,\cite{Serre02} with formula 
M(OH)(C$_8$H$_4$O$_4)$,
where is M is a trivalent species
such as Cr, Sc, Al, Ga or Fe.  
These structures consist of zigzag M-OH-M-OH$\dots$ chains,
crosslinked by 1,4-benzodicarboxylate O$_2$C-C$_6$H$_4$-CO$_2$ (bdc)
units (\fig{mil53x}). Each M is coordinated by two oxygens of OH units
and four carboxylate oxygens yielding octahedral oxygen
coordination.

These MIL-53 compounds exhibit a variety of topologically
equivalent structures with different volumes, but
generally include a narrow pore ($np$) structure 
and a large pore ($lp$) 
structure, both with formula M$_4$(OH)$_4$(bdc)$_4$ per conventional
unit cell, but with significantly different volumes.  In MIL-53(Al), 
the phase transition between $np$ and $lp$ forms can be reversibly achieved by 
cycling the temperature;\cite{Liu08}
the cell parameter corresponding to the direction of the
short axis of the lozenge pores was found to increase by
87~\% in the $np$-$lp$ transformation.  
By way of comparison, the strain variations achieved or predicted in functional 
``hard" materials such as 
(PbMg$_{1/3}$Nb$_{2/3}$O$_3$)$_{(1-x)}$-(PbTiO$_3$)$_{x}$\cite{Park97} or
BiFeO$_3$\cite{Dieguez11} are much smaller.
The large hysteresis\cite{Liu08} in the $np$-$lp$ phase transition of
MIL-53(Al) indicates that the transition is first-order.
Taking the transition temperature as the midrange of the hysteresis 
loop, the transition temperature $T_c$ is approximately 260 K; an estimate based on 
experimental sorption measurements places the transition at a somewhat lower
temperature of 203 K.\cite{Boutin10}

 For empty MIL-53(Cr), the $lp$ structure is thermodynamically
preferred at all temperatures.  In this system, a phase transition
to a $np$ structure has instead been observed in the case of
(1) sorption of a variety of sorbates; (2) pressure.
The hysteresis of the process in each case\cite{Serre07} indicates 
again that there is a transition barrier.  By fitting sorption isotherms, it 
was determined that the free energy difference between the $lp$ and $np$ forms 
of MIL-53(Cr) was only about 12 kJ mol$^{-1}$ of Cr$_4$(OH)$_4$(bdc)$_4$.\cite{Coudert08,
DevatourVinot09,Coombes09} An experiment that put the system under
hydrostatic pressure\cite{Beurroies10} came
up with a similar free energy difference.




 The phase transition of MIL-53(Al) was explained
by Walker et al.\cite{Walker10} in 2010.  
Van der Waals interactions stabilize the $np$
structure at low temperature, and vibrational 
entropy drives the structural transition to the 
$lp$ phase above $T_c$.  Density functional theory (DFT)
phonon calculations were used to quantify the vibrational
entropy. 
In that work, however,
the DFT energy and vibrational entropy were 
determined for only the $np$ and $lp$ structures.
However, to build an accurate picture of the $np$-$lp$ phase 
transition, including the hysteresis
and possible coexistence of $np$ and $lp$ 
phases,\cite{Triguero12} it is necessary to know 
the quantitative free energy landscape over 
the {\em full} volume range spanning the $np$ and $lp$ structures.  
This free-energy landscape of MIL-53 systems has previously 
been modeled in an {\it ad hoc} manner.\cite{Triguero11,Ghysels13}
This paper uses density functional total energy and
phonon linear response calculations to compute the
Helmholtz and Gibbs free energy in MIL-53(Cr)
as a function of temperature, pressure, and cell
volume, under the quasiharmonic approximation.
MIL-53(Cr) was chosen because of its relatively simple
phase transformation behavior and because it is
well-characterized experimentally.


 The thermodynamic calculations are performed within the 
quasiharmonic approximation.   In the quasiharmonic approximation,
the anharmonic lattice dynamics that leads to thermal expansion,
etc., is approximated by harmonic lattice dynamics where the
phonon frequencies are volume-dependent.   Suppose that one has a 
crystal where the rank-ordered frequencies ${\nu_{\mu} (V)}$ can be
determined for an arbitrarily large supercell (equivalently at 
arbitrary points in the Brillouin zone of the primitive cell).
The contribution of phonons to the thermodynamics is then given
well-known expressions.\cite{Maradudin71,vandeWalle02,Fultz10,Huang16}
Defining a dimensionless parameter $x_{\mu}(V,T) =
\frac{h \nu_{\mu}(V)}{k_B T}$, the molar internal energy as a function of
volume and temperature is given by
\begin{eqnarray}
\frac{U}{N}(V,T) = {\rm Lim}_{|a_{\rm min}|\rightarrow \infty}
\frac{1}{N} \bigl(U_0(V) + \nonumber \\
k_B T \sum_{\mu = 4}^{3 N_A}
[\frac{x_{\mu}(V,T)}{2} {\rm coth}(\frac {x_{\mu}(V,T)}{2})]\bigr),
\label{eq:inten}
\end{eqnarray}
the Helmholtz free energy by
\begin{eqnarray}
\frac{F}{N}(V,T) = {\rm Lim}_{|a_{\rm min}|\rightarrow \infty}
\frac{1}{N} \bigl(U_0(V) + \nonumber \\ 
k_B T \sum_{\mu = 4}^{3 N_A}
[\frac{x_{\mu}(V,T)}{2} + {\rm ln} (1 - e^{-x_{\mu}(V,T)})]\bigr),
\label{eq:helm}
\end{eqnarray}
and the Gibbs free energy is given by
$\frac{G}{N}(V,T) = \frac{F}{N}(V,T) + P V$.
$U_0(V)$ is the ground state energy neglecting zero-point vibrations, $N$
the number of moles and $N_A$ the number of atoms in the supercell, 
and the summation begins at $\mu = 4$ to avoid the weak singularity
due to the zero-frequency translational modes.

 First principles density functional theory calculations, as encoded in
the {\sc VASP} software~(\onlinecite{Kresse96,disclaim}), were used to 
compute  $U_0(V)$ and  ${\nu_{\mu} (V)}$ for a 152-atom supercell
of MIL-53(Cr), doubled along $c$ so as to make $a$, $b$, and $c$ similar in
magnitude for the $lp$ phase.  Two different sets of calculations were performed:
GGA calculations using the PBEsol functional\cite{Perdew08} and meta-GGA calculations
using the PBEsol+RTPSS\cite{Sun11} functionals.  These functionals were chosen because
we have had success with them in past studies of microporous 
materials.\cite{Cockayne12,Cockayne15}  For each level of DFT, the nonlocal 
van der Waals interactions were treated using three different approximations 
of Grimme et al.: DFT-D2,\cite{Grimme06} DFT-D3,\cite{Grimme10} 
and DFT-D3(BJ).\cite{Grimme11}
Anisotropic Hubbard parameters\cite{Liech95} were used for Cr and O atoms (GGA: U(Cr) = 4.0 eV,
J(Cr) = 0.5 eV; metaGGA: U(Cr) = 2.8 eV, J(Cr) = 0.5 eV; U(O) = 7.05 eV).
Spin polarized calculations were performed using the most-stable antiferromagnetic 
arrangement of charges on the Cr$^{3+}$ ions.
Further details of the DFT calculations are given in the Supplementary
Information (SI).

 Determination of $U_0(V)$ for each functional was done via straightforward 
fixed-volume relaxation for (primitive cell) increasing in 50~\AA$^3$ steps
from 650~\AA$^3$ to 1700~\AA$^3$.  The phonon frequencies for the 
152-atom supercell were calculated using ab initio linear response.  As 
this method converges toward exact second derivatives of the energy, it is more 
accurate than fitting frozen-phonon results.  Due to the large 
number of degrees of freedom, the phonon calculations are very expensive, and 
eventually only three calculations were used for the thermodynamics: 
V = 710~\AA$^3$, V = 1200~\AA$^3$, and V = 1506~\AA$^3$.  Linear response was only done 
using GGA and DFT-D2; the same phonon frequencies $\nu_{\mu}(V)$ were used for each 
functional in \eq{helm}; only the $U_0$ changed.  
Because the variation in volume between the $np$ and $lp$ phases is so large, one 
does not expect the conventional linear Gr\"uneisen approximation for $\nu_{\mu} (V)$
to apply.  Instead, we fit the phonon frequencies at intermediate volumes 
by fitting to the following physically-motivated expression:
\begin{equation}
\nu_{\mu}^2 (V) = \nu_{\mu \infty}^2 + C_1/V + C_2/V^2 .
\label{eq:phofit}
\end{equation}
The coefficients in~\eq{phofit}  were determined by fitting the results for the three
frequencies calculated.  If $\nu_{\mu \infty}^2$ in the fit was less than
zero, it was set to zero and the fit recalculated.
Due to computational limitations, it is not possible to calculate larger 
supercells for use in~\eq{helm}.  Instead, the contribution of optical phonons to 
the thermodynamics was approximated by the phonon spectra calculated for 
the single 152-atom supercell.  The contribution of acoustic phonons to the 
thermodynamics was approximated by numerical integration of estimated acoustic 
frequencies over the first Brillouin zone.  Further details are given in the 
Supplementary Information.

\begin{figure}
\includegraphics[width=42mm]{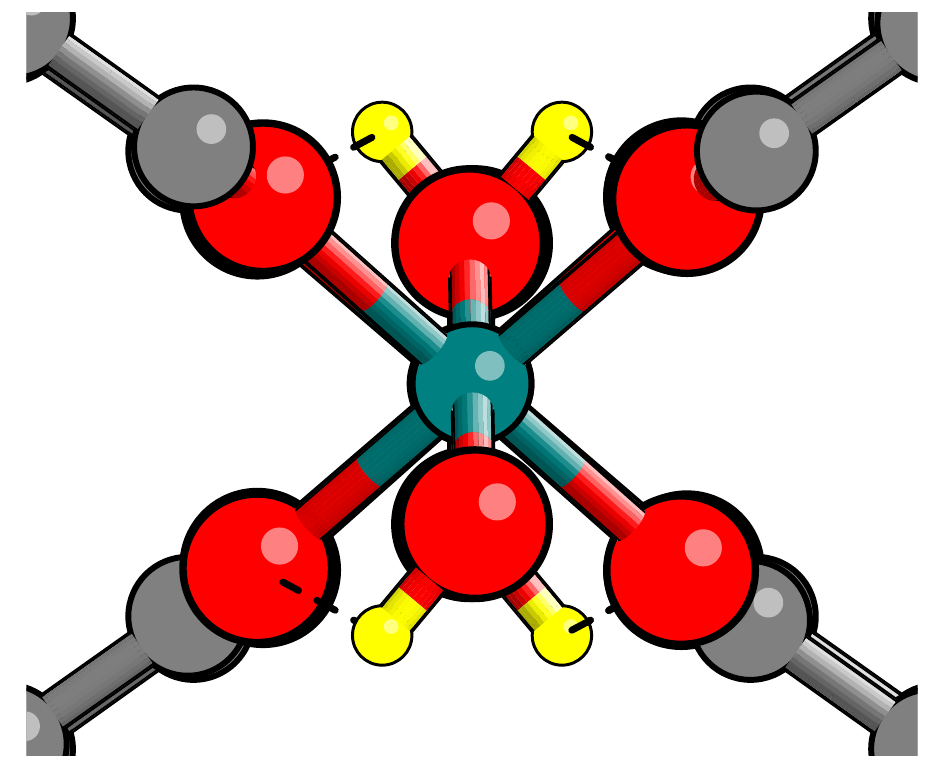}
\caption{Local geometry of MIL-53(Cr) (lp) after DFT relaxation of
``H flopping" mode.  Each H relaxes to sit approximately 2.4 \AA~from
each of a pair of oxygens (dashed lines); the O are superposed 
from this vantage point.}
\label{fig:flop}
\end{figure}

 First, the phonons were calculated for the $np$ and $lp$
structures.  All modes were stable for the $np$ structure.
For the $lp$ structure, instabilities were found.  The
most unstable modes, for both the force-constant and dynamical
matrices,  were hydrogen ``flopping" modes in which
the H in each hydroxyl group move in the $\pm x$ direction
so as to decrease the distance to a pair of carboxylate 
oxygens (\fig{flop}).
Fully relaxing this mode maintains orthorhombic symmetry,
the 152-atom cell is now a primitive cell.

The structure obtained upon relaxation of the flopping instability
was taken as the reference $lp$ structure.  To obtain the initial
structure for the fixed volume relaxations used to determine $U_0(V)$,
the ionic coordinates were interpolated (or extrapolated) from the initial
$np$ and $lp$ structures.

\begin{figure}
\includegraphics[width=85mm]{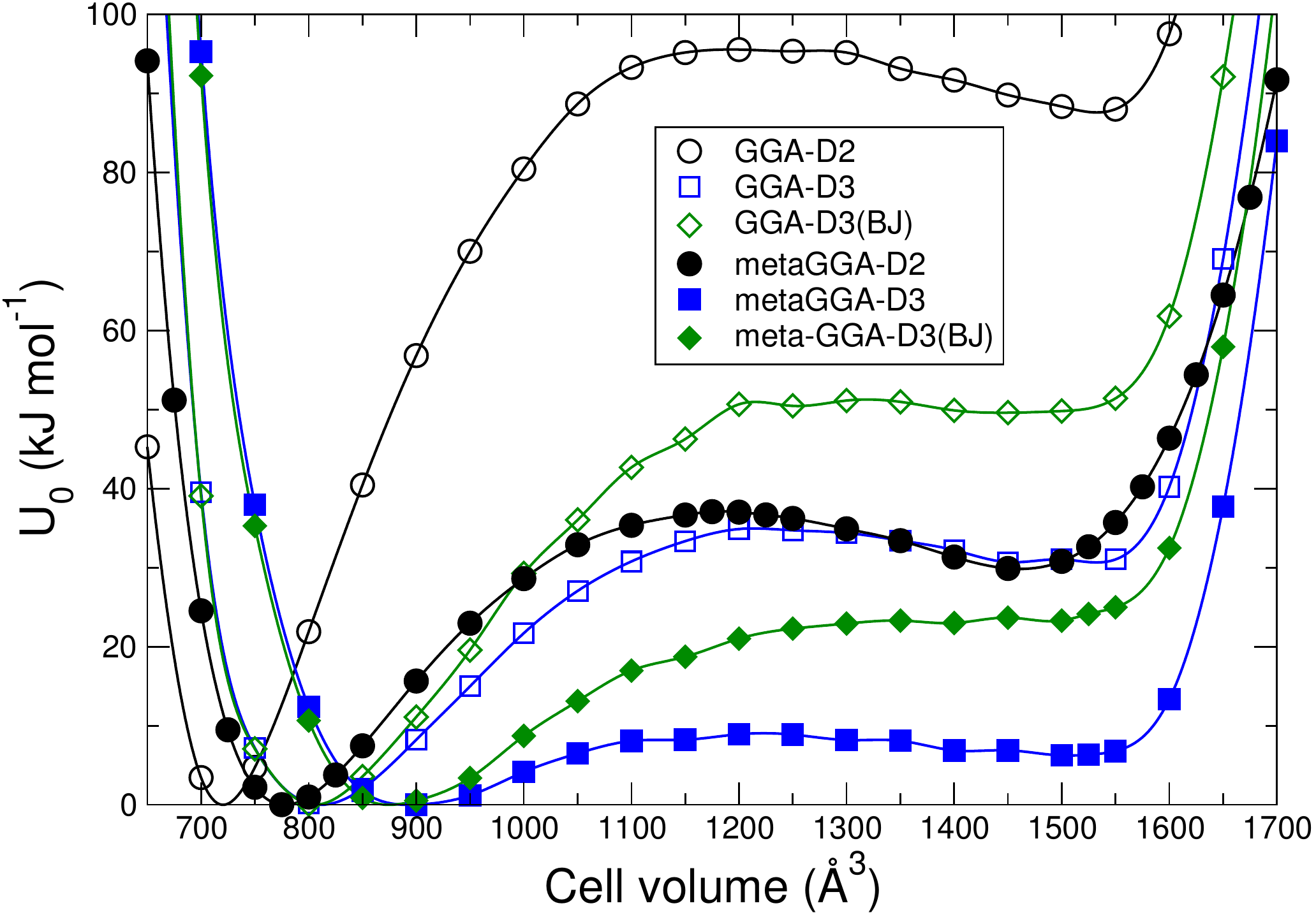}
\caption{Calculated DFT energy for MIL-53(Cr) at 0 K as a function of
volume for different density functionals, neglecting zero-point motion. Each curve is 
scaled so that its minimum is zero.}
\label{fig:uo}
\end{figure}

\begin{figure}
\includegraphics[width=85mm]{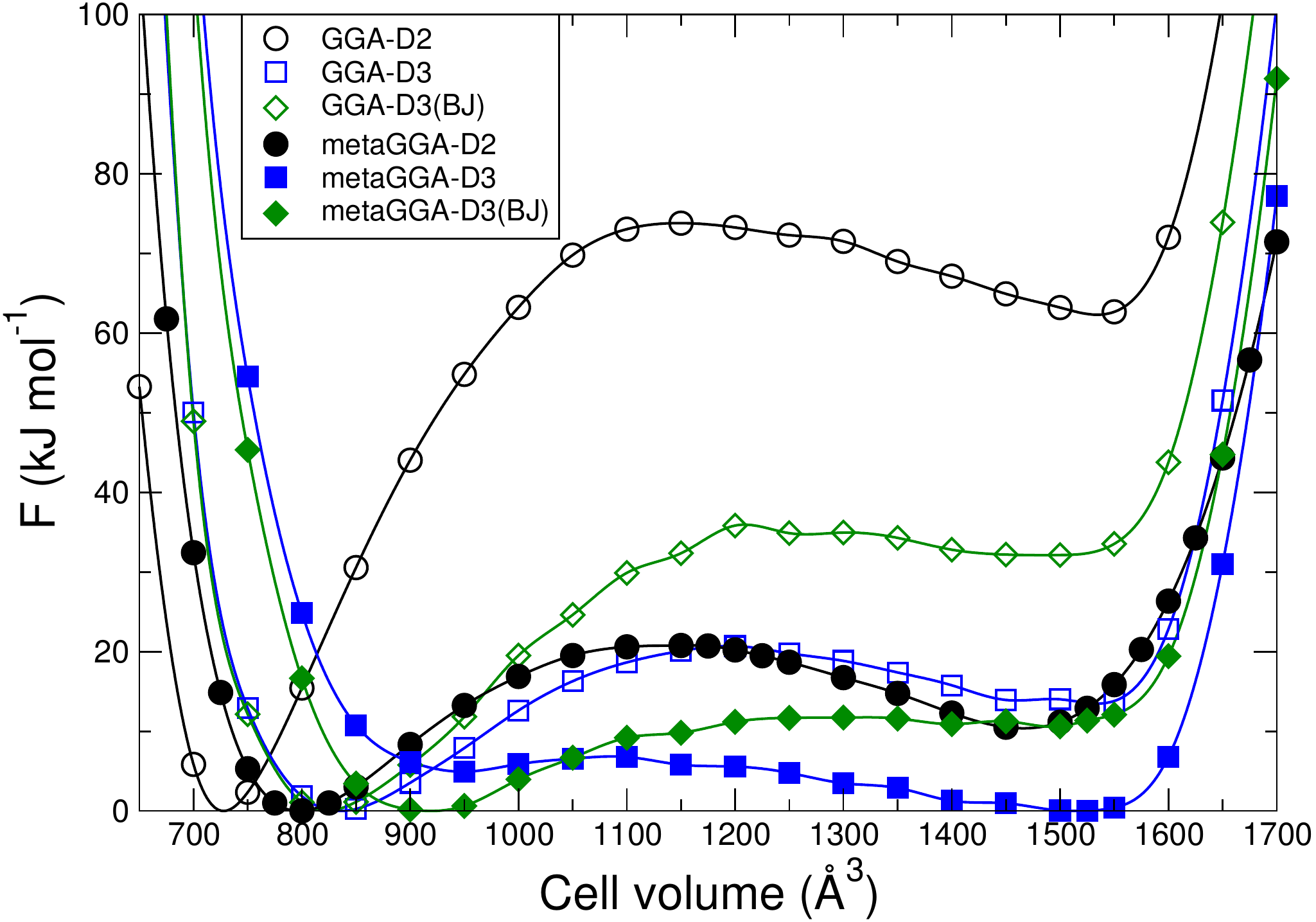}
\caption{Calculated Helmholtz free energy for MIL-53(Cr) at 293 K as a function of
volume for different density functionals. Each curve is scaled so that its minimum 
is zero. The effect of atmospheric pressure of about 0.1 MPa is negligible on this 
scale.}
\label{fig:helm}
\end{figure}

The $U_0(V)$ determined for the various density functionals are shown in 
\fig{uo}. The $F(V)$ for T = 293 K are shown in \fig{helm}.
For every plot in \fig{helm}, there are two minima in the free
energy, corresponding to $lp$ and $np$ structures.  The effect of
phonon entropy is to reduce the free energy of the $lp$ structure with
respect to the $np$ structure, as expected.  Calculations show that the 
free energies for temperatures up to 500 K and pressures between 
-30 MPa and 30 MPa maintain two minima for all density functionals tested.

\tab{break} summarizes and compares the results for the
different functionals used.  The volumes at which the minima for $U_0$ occur
are given by $V_{0np}$ and $V_{0lp}$.  The locations of the  minima in
$F$ at room temperature (RT; 293K) are given by $V_{np}(RT)$ and
$V_{lp} (RT)$.  The calculated difference in $F$ between the $np$ and $lp$ minima
is $\Delta{F}(RT) =  F_{lp}(RT) - F_{np}(RT)$.
The critical pressure $P_c$ is where the calculated Gibbs free energy of the $np$
and $lp$ phases becomes equal at T =  293 K.  $G_{b}(RT;P_c)$ is the
calculated free energy barrier between the phases at this pressure.


\begin{table}
\caption{Calculated structural and thermodynamic results for MIL-53(Cr) for different choices of
the density functional.  See text for explanation of the column headings.  
$V$ are in~\AA$^3$; $F$ and $G$ in kJ mol$^{-1}$ 
(1 mole = 1 mole of Cr$_4$(OH)$_4$(bdc)$_4$); $P$ in MPa.  RT is
room temperature, or 293 K.}
\resizebox{\columnwidth}{!}{%
 \begin{tabular}{ccccccccc}
  \hline
xc      & vdW     &  $V_{0np}$ & $V_{0lp}$ & $V_{np} (RT)$ & $V_{lp} (RT)$ & $\Delta{F}$ & $P_c$ & $G_{b}(RT;P_c$) \\
GGA     & D2     &   720  &  1533 &  728   &   1534  &  +62.3  &  -127.0 &  45.6  \\
GGA     & D3     &   811  &  1532 &  835   &   1534  &  +13.5  &   -32.0 &  14.1  \\
GGA     & D3(BJ) &   806  &  1443 &  822   &   1483  &  +32.1  &   -76.8 &  18.7  \\
metaGGA & D2     &   778  &  1461 &  798   &   1466  &  +10.4  &   -25.6 &  16.0  \\
metaGGA & D3     &   892  &  1505 &  948   &   1512  &   -4.9  &    14.6 &   3.2  \\
metaGGA & D3(BJ) &   875  &  1393 &  919   &   1493  &  +10.6  &   -30.7 &   6.0  \\
\hline
\label{tab:break}
\end{tabular} 
}
\end{table}

 Substantial differences are seen depending on what density functional is used.
The general trend is for the GGA functionals and the D2 vdW term to give lower
$V_{np}$ and higher $\Delta{F}$ than the metaGGA functionals and D3 or D3(BJ)
choices for the vdW interaction.  Which functional gives the best agreement with
experiment?  The experimental unit cell volume of the $lp$ phase of MIL-53(Cr) is
1486~\AA$^3$.(Ref.~\onlinecite{Llewellyn08})  The volume of the $np$ phase formed upon
sorption of H$_2$O is 1012 ~\AA$^3$,(Ref.~\onlinecite{Llewellyn08}) but this cannot
be directly compared with the calculation for the empty cell reported here.  
As the $np$ phase of MIL-53(Cr) is thermodynamically unstable experimentally, we
take the experimental volume\cite{Liu08,Nanthamathee15} of MIL-53(Al) $np$, 864~\AA$^3$, 
and estimate that the volume of MIL-53(Cr) should be about 900~\AA$^3$ due to the larger ionic
radius of Cr$^{3+}$.  The best agreement with experiment for the lattice parameters is
for the metaGGA-D3(BJ) parameterization, while the second best is for metaGGA-D3.
On the other hand, the relative stability of the $lp$ phase found experimentally,
$\Delta{F} \approx$ -12.0 kJ mol$^{-1}$ is underestimated by {\em all} the functionals
chosen.  The metaGGA-D3 calculation is best in this regard, as it is the only
calculation to yield a negative $\Delta{F}$.  All of the metaGGA calculations perform
better than GGA in predicting the relative phase stability.  As the metaGGA-D3 and
metaGGA-D3(BJ) have the best agreement with experiment, their low values 
of the transition barrier $G_b$, 3.2 to 6.0 kJ mol$^{-1}$ should be considered most
reliable.

 It is interesting to put the comparative results in context of previous studies.
In MIL-53, it has previously been found that the D2 vdW overbinds the 
$np$ phase;\cite{Haigis14} this work confirms that result.  Benchmarking 
the performance of DFT calculations
is currently receiving a great deal of attention\cite{Kirklin15,Lejaeghere16,Tran16}.  
In Ref.~\onlinecite{Tran16}, over sixty different density functionals are
compared.  Although the RTPSS functional is not tested, the related metaGGA functional 
TPSS-D3 gives good results for graphite, which suggests that these parameterizations
may work well for MIL-53, where the $np$ phase has benzyl rings of carbon approaching
each other.  Further work is needed to make a full comparison among methods because the
current work: (1) includes Hubbard U and J parameters; (2) needs a vdW functional that
reproduces the vdW interactions correctly over a wide range of structural distortion, 
not merely at one equilibrium point.

 The metaGGA-D3 calculation predicts that the $lp$ phase of MIL-53(Cr) is stable
at room temperature, in agreement with experiment.  Interestingly, it predicts
a transition to the $np$ phase below T = 160 K, similar to what actually occurs for MIL-53(Al).
The estimated change in $\Delta{F}$ with temperature is about 
-0.036 kJ mol$^{-1}$ K$^{-1}$.  Applying this to the experimental $\Delta{F} \approx$
-12.0 kJ mol$^{-1}$,  the $lp$ phase is expected to remain stable down to T =  0 K,
albeit with a free energy advantage of less than 2 kJ mol$^{-1}$.

\begin{figure}
\includegraphics[width=85mm]{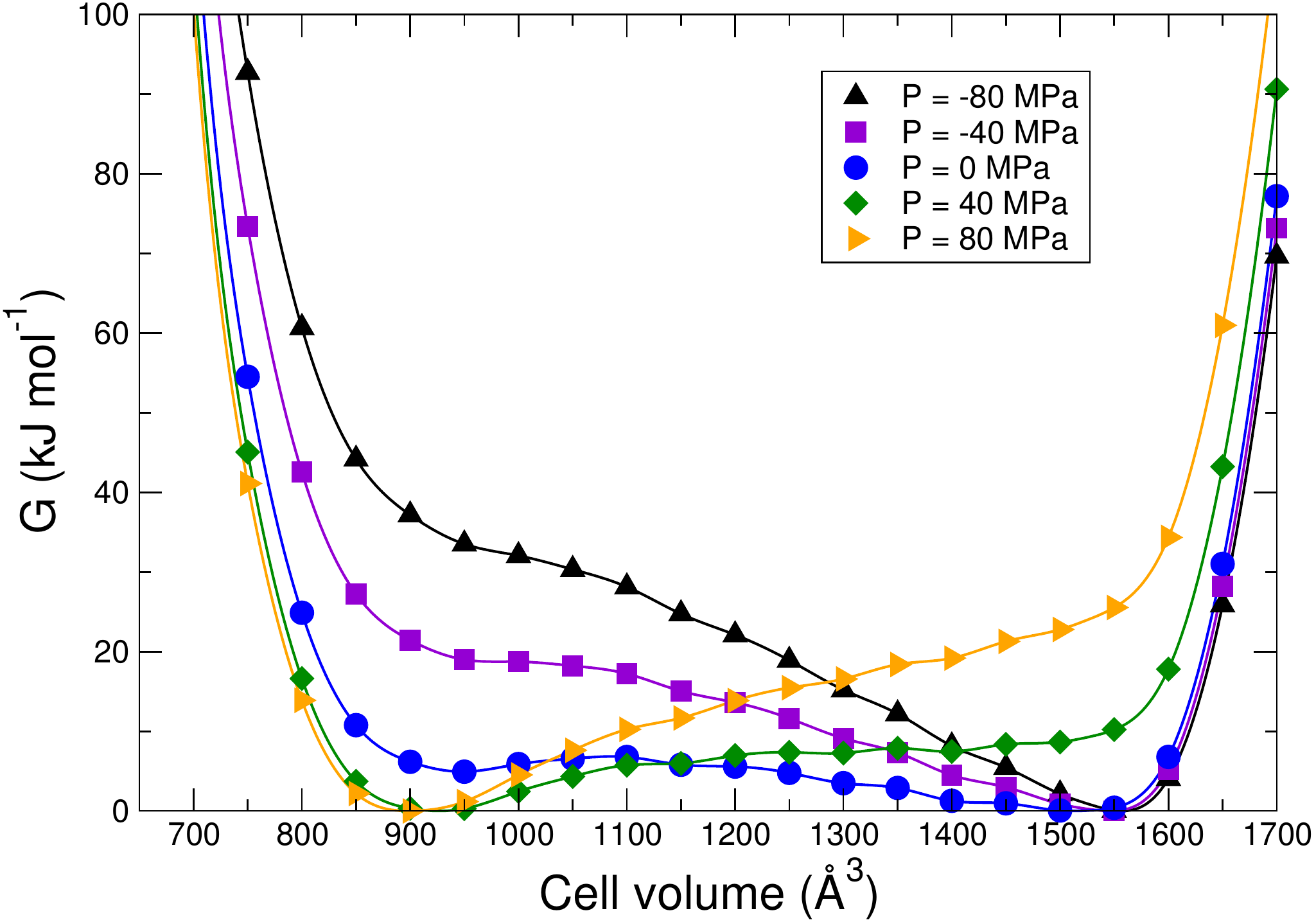}
\caption{Calculated Gibbs free energy for MIL-53(Cr) at 293 K as a function of
volume and pressure for the metaGGA-D3 density functional.  Each curve is
scaled so that its minimum is zero.}
\label{fig:gibb}
\end{figure}

 The shallowness of the free energy profile suggests that sufficiently large
positive or negative pressure would drive the Gibbs free energy
G(V, T = 293 K) into a regime where it has only one minimum corresponding to
either a $np$ or a $lp$ structure.  In \fig{gibb}, we show G(V, T = 293 K) for 
various pressures -80 MPa to 80 MPa, using the metaGGA-D3 results.
At pressures above about 60 MPa, there is a unique minimum at the $np$ phase;
below about -40 MPa, there is one minimum at the $lp$ phase.  If the zero in
pressure is shifted to correct for the error in the  metaGGA-D3 $\Delta{F}$ with
respect to experiment, the predicted pressures are shifted to about 80 MPa and -20 MPa,
respectively.  Of course the prediction of the pressures at which the free energy converts to
a single minimum only sets an upper bound on the width of the pressure hysteresis loop;
in practice, fluctuations will cause the transitions to occur at less extreme pressures.
With this is mind, experimental transition pressures for the hysteresis loop 
of roughly 50 MPa and 20 MPa for MIL-53(Cr)\cite{Rodriguez16} are
consistent with the DFT results.  Note that negative pressures do have physical 
relevance in microporous materials in the case of sorption- the effective solvation pressure 
can be either positive or negative depending on the sorbate concentration.\cite{Ravikovitch06}



\begin{figure}
\includegraphics[width=85mm]{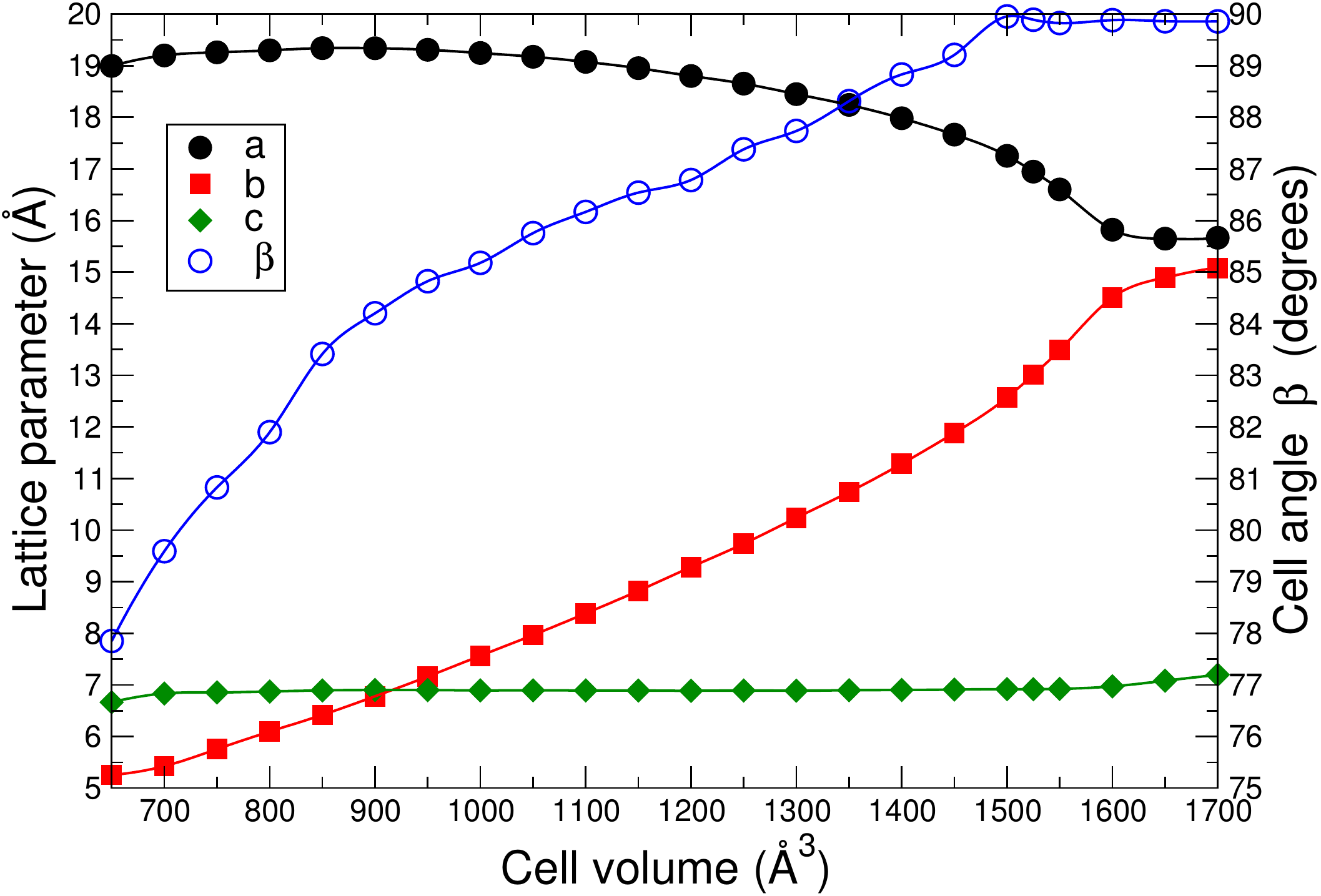}
\caption{Calculated MIL-53(Cr) lattice parameters and cell angle $\beta$ versus volume.}
\label{fig:cell}
\end{figure}

In \fig{cell}, the crystallographic data for the DFT metaGGA-D2 structural
relaxations are shown.   The lattice parameters are scaled to the volume of the
conventional unit cells.  To make the orthorhombic-monoclinic transition
clear, the monoclinic cell parameters $a$ and $\beta$ are for an unconventional
body-center monoclinic setting.
The orthorhombic-monoclinic transition occurs at $V \sim 1500$~\AA$^3$,
intriguingly close to the experimental cell volume.
In addition to the structural transitions, there are three regimes in
the behavior of the lattice constants: (1) below about 850~\AA$^3$, $a$ $b$
and $c$ all increase with volume; (2) between about 850~\AA$^3$ and
1650 \AA$^3$, $a$ decreases with volume $b$ increases with volume, and 
$c$ is nearly flat as the structure flexes; (3) above about 1650 \AA$^3$, all
lattice parameters increase again.   The crossover between regimes
(2) and (3) does not occur at the same volume as the 
monoclinic-orthorhombic transition.  To a first approximation, the free energy 
is nearly flat in regime (2) and increases rapidly above and below this range.
The three regimes agree qualitatively with those seen in a recent 
experiment on the related material MIL-53(Al) under pressure.\cite{SerraCrespo15}


 To summarize, we used density functional theory
total energy and linear response phonon calculations
to compute the free energy profile of MIL-53(Cr)
under the quasiharmonic approximation.  
The density functionals that best match the experimental
results give remarkably flat free energy profiles,
with a transition barrier of only about a 3 to 6 kJ mol$^{-1}$
between the the narrow pore and large pore phases.


 I thank Laura Espinal, Kevin F. Garrity,
and Winnie Wong-Ng for helpful discussions.

This paper was published as J. Phys. Chem. C 2017,
{\bf 121}, 4312-4317 (DOI:10.1021/acs.jpcc.6b11692).
The Supporting Information is available free of charge on the
ACS Publications website at the DOI given immediately
above.


\begin{thebibliography}{42}
\bibitem [{dis()}]{disclaim}%
  \BibitemOpen
  \href@noop {} {}\bibinfo {note} {Certain commercial software is identified in
  this paper to adequately describe the methodology used. Such identification
  does not imply recommendation or endorsement by the National Institute of
  Standards and Technology, nor does it imply that the software identified is
  necessarily the best available for the purpose.}\BibitemShut {Stop}%
\end{thebibliography}

%
\end{document}